\documentclass[proof]{pasj00}
\draft

\begin{document}
\SetRunningHead{Author(s) in page-head}{Running Head}
\Received{2010/11/21}
\Accepted{2010/12/23}

\title{Size Distribution of Main-Belt Asteroids with High Inclination
  \thanks{Based on data collected at Subaru Telescope, which is operated by the National
  Astronomical Observatory of Japan (NAOJ).}}

\author{Tsuyoshi \textsc{TERAI} and Yoichi \textsc{ITOH}}
\affil{Graduate School of Science, Kobe University, 1-1 Rokkodai, Nada-ku, Kobe 657-8501, Japan}
\email{terai@stu.kobe-u.ac.jp}

%


\maketitle

\begin{abstract}
We investigated the size distribution of high-inclination main-belt asteroids (MBAs) to
explore asteroid collisional evolution under hypervelocity collisions of around 10 km s$^{-1}$.
We performed a wide-field survey for high-inclination sub-km MBAs using the 8.2-m Subaru
Telescope with the Subaru Prime Focus Camera (Suprime-Cam).
Suprime-Cam archival data were also used.
A total of 616 MBA candidates were detected in an area of 9.0 deg$^2$ with a limiting magnitude of
24.0 mag in the SDSS $r$ filter.
Most of candidate diameters were estimated to be smaller than 1 km.
We found a scarcity of sub-km MBAs with high inclination.
Cumulative size distributions (CSDs) were constructed using Subaru data and published
asteroid catalogs.
The power-law indexes of the CSDs were 2.17 $\pm$ 0.02 for low-inclination ($< 15^{\circ}$) MBAs
and 2.02 $\pm$ 0.03 for high-inclination ($> 15^{\circ}$) MBAs in the 0.7--50 km diameter range.
The high-inclination MBAs had a shallower CSD.
We also found that the CSD of S-like MBAs had a small slope with high inclination, whereas the slope
did not vary with inclination in the C-like group.
The most probable cause of the shallow CSD of the high-inclination S-like MBAs is the large
power-law index in the diameter--impact strength curve in hypervelocity collisions.
The collisional evolution of MBAs may have advanced with oligopolistic survival
during the dynamical excitation phase in the final stage of planet formation.
\end{abstract}

\section{Introduction}

The size distribution of main-belt asteroids (MBAs) traces their collisional history since the
formation of primitive planetesimals.
Previous survey observations for small solar system bodies have revealed the size distribution of
MBAs down to sub-km size (Ivezi\'c et al. 2001, Yoshida \& Nakamura 2007, Gladman et al. 2009).
Understanding the collisional processes of MBAs provides insight into the initial size
distribution and collisional and dynamical evolution of planetesimals in the inner protoplanetary
disk.

Several models for collisional and orbital evolution of MBAs suggest a dynamical excitation
event at the final stage of the planet formation \citep{P02}.
This event produced several features seen in the current main belt, such as a high
eccentricity/inclination asteroid population and the radial mixing of taxonomic classes.
The dynamical excitation also ejected most asteroids out of the main belt, depleting more than 99\%
of the primordial mass \citep{Bo05}.
The relative velocities among asteroids were elevated during this phase.
In the main belt, asteroid collisions mostly occurred at higher velocities than the present average
velocity of $\sim$4 km s$^{-1}$ \citep{V98}.
The relative velocity between a body in the remnant main belt population and one in the ejected
population from the main belt zone exceeded 10 km s$^{-1}$ \citep{Bo05}.
However, the fragment size distribution and ejecta velocities of such high-velocity collisions are
still unclear; reproduction by laboratory experiments is difficult \citep{K10}.

In our approach to this problem, we focus on MBAs with high inclination.
The mean relative velocity of highly-inclined MBAs at collisions with other MBAs can be as large as
$\sim$10 km s$^{-1}$ \citep{G06}.
Comparing the size distributions between low-inclination (low-$i$) and high-inclination
(high-$i$) MBAs could provide meaningful insight into disruptions at high velocity. 
For investigating high-$i$ MBAs, several published asteroid catalogs are available, but
their detection limits are no smaller than kilometer size.
We performed observations for sub-km high-$i$ MBAs to obtain their size distribution.

\section{Observations}

\subsection{Strategies}

The inclination distribution of the known MBA population drops steeply at $i \sim$$15^{\circ}$,
where $i$ is the inclination.
The scarcity of high-$i$ MBAs is due not only to the small numbers but also to an observational
bias.
Most previous MBA surveys have searched intensively around the ecliptic, but high-$i$ MBAs only
pass through the ecliptic plane briefly, spending most of their time in high ecliptic latitudes.
High-$i$ bodies are more likely to be found at $\beta_{\rm h} \sim i$ where $\beta_{\rm h}$ is the
heliocentric ecliptic latitude than at low ecliptic latitudes ($\beta_{\rm h} < i$).
We surveyed high ecliptic latitudes to detect high-$i$ MBAs.

Ivezi\'c et al. (2001) showed that the surface density of asteroids in
$\beta \sim \pm 20^{\circ}$ is only $\sim$10 $\%$ of that in $\beta \sim 0^{\circ}$
, where $\beta$ is the geocentric ecliptic latitude (a factor of 1.5--2 smaller than
$\beta_{\rm h}$ at opposition for MBAs).
To find a number of high-$i$ MBAs, a wide-field survey is indispensable.
We performed a wide-field survey for faint MBAs using data derived from the 8.2-m Subaru
Telescope.
In addition, efficient observation is necessary to collect large amounts of data within a limited
amount of time.
We adopted a less usual procedure, observing each field only two times with an interval of about
20 minutes.
Terai et al. (2007) developed a new reduction technique to find asteroids efficiently in
such two-visit observation data sets.

\subsection{Observations}

Observations were conducted on June 3 2008 (UT) with the Subaru Telescope at the summit of
Mauna Kea in Hawaii.
We used the Subaru Prime Focus Camera (Suprime-Cam), a mosaic camera of ten CCD chips.
It covers a 34$^{\prime} \times 27^{\prime}$ field-of-view with a pixel scale of
0.20$^{\prime\prime}$ and has a high sensitivity of $R_{c}$ = 25.6 mag in 5-minutes integration
for a point source \citep{M02}.
The observational fields are near R.A.(J2000) = \timeform{16h59m}, Dec.(J2000) = \timeform{+02D26'}
corresponding to $\beta \sim 25^{\circ}$ at opposition.
We selected 22 fields with no bright stars, two of which were located on the Sloan Digital
Sky Survey (SDSS) fields.
Several stars cataloged in the SDSS
database\footnote{Sloan Digital Sky Survey$\langle$http://www.sdss.org/$\rangle$.} were used for
photometric calibration.
All data were acquired with the SDSS $r$-band filter \citep{F96}.
The exposure time was 4 minutes.
Each field was taken twice with an interval of about 20 minutes.
The total observed area of the sky was $\sim$5.6 deg$^{2}$ in 4 hours.
The typical seeing size was 0.6 arcsec.
The detection limit range was $r$ = 23.8--24.2 mag for moving objects whose motion corresponded to
MBAs (see Section 4.2).
We used data with a limiting magnitude of $r$ = 24.0 mag.
The available data includes 14 fields, 3.6 deg$^{2}$ (Table 1).

\subsection{Archival Data}

We also used archival Suprime-Cam data obtained from the Subaru-Mitaka-Okayama-Kiso
Archive, SMOKA\footnote{The SMOKA is operated by the Astronomy Data Center, NAOJ$ $.}
\citep{Ba02}, consisting of $R_{c}$-band data with a detection limit of $R_{c} \ge$ 23.8 mag for
moving objects (Section 4.2).
The selected data were taken where the ecliptic latitudes with respect to opposition were within
$\pm 15^{\circ}$, allowing reasonably accurate orbital estimations from the short arcs.
We avoided star crowding regions around the galactic plane.
Each field was taken at least twice, at intervals of 5 minutes to 1 hour.
We searched a total of 28 fields (Table 2).

\section{Data Reduction}

The images were processed chip-by-chip using IRAF produced by the National Optical Astronomy
Observatories (NOAO).
The raw data were reduced with standard processes consisting of bias subtraction, flat-fielding,
sky subtraction, distortion correction, and positional correction.
We also used SDFRED, the software package for Suprime-Cam data reduction (Yagi et al. 2002,
Ouchi et al. 2004) to subtract the sky background and correct for geometric distortion.

Detecting a moving object from two-visit data is a three-step procedure: (i) masking fixed stars
and galaxies in the field, (ii) detecting the remaining sources from the processed images, and
(iii) removing non-object sources such as cosmic rays.
First, the fixed objects were masked with the image-processing technique developed by Terai et al.
(2007).
Two types of composite images were made from raw images: one excluding smaller values of each pixel
(``OR" image) and one excluding larger values of each pixel (``AND" image).
The OR image included all objects in either of the raw images, whereas the AND image consisted
of only fixed objects.
The AND image was used to make a mask image whose pixel values were replaced with 0 for
light sources and 1 for background.
This was multiplied by the OR image to mask the fixed objects.
A moving object was recognized as a pair of light sources in the final processed image
(see Fig. 1).

The remaining sources were detected with the SExtractor software \citep{BA96}.
The detection threshold was set to 2$\sigma$, where $\sigma$ is the root-mean-square value of the
background.
But a large number of false sources were also identified.
Most were discriminated using criteria such a sharpness of the point spread
function (PSF), velocity of motion, and direction of motion.
An object pair that fulfils all of these criteria is considered a moving object candidate.
Finally, each candidate was inspected visually to determine whether or not it was in fact a moving
object.

\section{Results}

\subsection{Detection}

We searched for moving objects in our observation data of 14 fields and the archival data of 28
fields.
In the total area of 9.0 deg$^{2}$, 656 moving objects were detected.
The apparent motion of each object was measured from the shift in position between two images.
Figure 2 shows the apparent motion distribution of the detected objects.
Most of the objects have motion of $\sim$30--40 arcsec hr$^{-1}$ along the ecliptic longitude,
corresponding to MBAs.
The smallest motion rate was 3 arcsec hr$^{-1}$, equivalent to a trans-Neptunian object, and the
largest was 107 arcsec hr$^{-1}$.
Highly inclined asteroids exhibited large motions in both the ecliptic longitude and latitude.

\subsection{Photometry}

The flux of each moving object was measured from the total intensity of pixels within a
non-circular aperture.
We used a capsule-shaped aperture given by a trail of a circle with the triple-FWHM (full
width at half maximum) diameter in length of the motion during the exposure time.
The apparent magnitude of an asteroid was estimated from the mean flux in the two images.
Flux calibration was conducted using several stars listed in the SDSS database for our
observation $r$-band data and the Landolt standard stars for the archival $R_c$-band data.
The limiting magnitude for moving objects was evaluated by artificial trails added to the raw
images.
They were made by imbricate pseudo point sources with the same FWHM of stars in the image.
We used 100 artificial trails with brightness steps of 0.1 mag.
We regarded the brightness corresponding to detection probability of 80 \% as the limiting
magnitude, which for our observational data was $r$ = 24.0 mag.
This was transformed to 23.8 mag for the $R_c$-band \citep{F96}.
We used the archival data when its limiting magnitude was larger than $R_c$ = 23.8 mag.
Within the limiting magnitude, the photometric error was less than 0.1 mag.

\subsection{Orbital Estimations}

To determine the orbit of an asteroid, we need at least three observations at different epochs.
In this study, we could not obtain precise orbits for the detected moving objects because we
observed a field only twice in a short arc of several dozen minutes.
Instead, we assumed a circular orbit.
This enabled us to estimate the semi-major axis and inclination of a moving object from
its sky motion (Bowell et al. 1990, Nakamura \& Yoshida 2002).
These elements were calculated using the equations for geocentric motion rate of a moving object
presented by Jedicke (1996) with an eccentricity of zero.
An object was excluded as an inaccurate orbit if its inclination was calculated to be larger than
40$^{\circ}$ (9 objects).

We examined the estimation accuracy using the Monte Carlo simulations in Nakamura and Yoshida
(2002).
We generated 10,000 pseudo-MBAs with randomly distributed orbital elements.
The ranges of their semi-major axes ($a$), eccentricities ($e$), and inclinations ($i$) were
$a=$2.1--3.3 AU, $e=$0.0--0.4, and $i=$0$^{\circ}$--30$^{\circ}$, respectively.
The eccentricity of each orbit was given with a probability based on the actual eccentricity
distribution of known MBAs.
The true anomaly was set with a probability following Kepler's second law.
The apparent velocity of motion was calculated for a hypothetical observation field from the
formulae in Jedicke (1996).
Then, we estimated the semi-major axis and inclination assuming a circular orbit from the velocity
of motion.

The accuracy of each element was represented by the standard deviations of differences between the
estimated value and the given value.
If the separation from opposition in ecliptic longitude was less than 15$^{\circ}$,
the semi-major axis error was $\sim$0.1--0.15 AU for a semi-major axis range of 2.5--3.3 AU.
The error was larger than 0.15 AU for semi-major axes between 2.0--2.5 AU.
The inclination accuracy depends strongly on inclination.
Nakamura and Yoshida (2002) mentioned the degradation in inclination accuracy for $i > 10^{\circ}$
(the random error of inclination is more than 5$^{\circ}$).
However, accuracy is improved in high ecliptic latitudes.
The inclination error is 3$^{\circ}$--4$^{\circ}$ at $\beta \sim 25^{\circ}$ where most high-$i$
MBAs were detected.
This is small enough to obtain precise size distributions for low-$i$ and high-$i$ MBAs.

The semi-major axis distribution of the detected moving objects is shown in Figure 3(a).
The distribution is concentrated between 2.0$-$3.3 AU, similar to what is known for in the
main belt.
We regarded 616 objects within this zone as the MBA sample.
The semi-major axis distribution of the sample with $a >$ 2.5 AU was similar to the actual
population, which had a local minimum at 2.8 AU and an outer edge at 3.3 AU, corresponding to the
5:2 and 2:1 mean-motion resonances (MMR) with Jupiter, respectively.
In contrast, the distribution in $a <$ 2.5 AU did not appear to replicate the known distribution
well.

The inclination distribution is shown in Figure 3(b).
The two local peaks at $\sim$10$^{\circ}$ and $\sim$16$^{\circ}$ are quite different from the
known distribution in the main belt.
This is because many of the surveyed fields are situated at geocentric ecliptic latitudes ($\beta$)
of 11$^{\circ}$, 17$^{\circ}$, and 25$^{\circ}$, corresponding to $\beta_{\rm h} \sim$ 7$^{\circ}$,
11$^{\circ}$, and 16$^{\circ}$, respectively.
This means that the highest probability MBA detections occur at inclinations of around 7$^{\circ}$,
11$^{\circ}$, and 16$^{\circ}$ (see Section 2.1).

The distribution of semi-major axis vs. inclination is plotted in Figure 4.
Most bodies were located between the secular resonance $\nu_6$ and the 2:1 MMR, which agrees with
the distribution of known MBAs.

\subsection{Diameter Estimations}

The absolute magnitude of an asteroid is defined as the brightness observed at a hypothetical
location of $r$ = 1 AU and $\Delta$ = 1 AU with the Sun, the asteroid, and Earth in a straight
line.
It is given by
\begin{equation}
H = m - 5\log(r\cdot\Delta) - P(\alpha),
\end{equation}
where $H$ is the absolute magnitude and $m$ is the apparent magnitude in a given wavelength band.
The apparent magnitude is reduced by the phase angle $\alpha$, the Sun-asteroid-Earth angle
following the phase function given by
\begin{equation}
P(\alpha) = (1-G)\Phi_1(\alpha)+G\Phi_2(\alpha),
\end{equation}
where $G$ represents the gradient of the phase function \citep{B89}.
We used $G$ = 0.15, which is generally used for asteroids with unknown taxonomic types.
$\Phi_1$($\alpha$) and $\Phi_2$($\alpha$) are given by
\begin{equation}
\Phi_i = \left[ -A_i \left( \tan \frac{1}{2} \alpha \right)^{B_i} \right] \quad
\left( i = 1,2 \right),
\end{equation}
where $A_1$ = 3.33, $A_2$ = 1.87, $B_1$ = 0.63, and $B_2$ = 1.22 \citep{B89}.
Although these are $V$-band values, we applied them to $r$- and $R_c$-band data.

The relationship between absolute magnitude and diameter ($D$) is described by
\begin{equation}
H = m_\odot - 5 \log \frac{\sqrt{p} D/2}{\mbox{(1 AU)}},
\end{equation}
where $m_\odot$ is the magnitude of the Sun, -27.29 mag in the $R_c$-band \citep{DL00}.
This corresponds to -27.07 mag in the $r$-band, derived from the color transformation equations in
Fukugita et al. (1996).
Albedo is represented by $p$, which cannot be determined in this survey.
Therefore, we used typical albedo values.

Asteroids have different albedos among the taxonomic groups.
In photometric observation studies, asteroids are divided into two major groups, S-like and C-like
asteroids \citep{YN07}.
We estimated the mean albedos of S-like and C-like MBAs from the albedo catalog presented
by the IRAS Minor Planet Survey and Supplemental IRAS Minor Planet
Survey\footnote{Tedesco, E. F., P. V. Noah, M. Noah, and S. D. Price. IRAS Minor Planet Survey.
IRAS-A-FPA-3-RDR-IMPS-V6.0. NASA Planetary Data System, 2004.
$\langle$http://sbn.psi.edu/pds/resource/imps.html$\rangle$.}
 (IMPS and SIMPS, respectively; Tedesco et al. 2002).
The cataloged asteroids were classified into S-like and C-like groups according to the
spectral database presented by the second phase of the Small Main-belt Asteroid Spectroscopic
Survey\footnote{Bus, S. and Binzel, R. P., Small Main-belt Asteroid Spectroscopic Survey, Phase II.
EAR-A-I0028-4-SBN0001/SMASSII-V1.0. NASA Planetary Data System,2003.
$\langle$http://sbn.psi.edu/pds/resource/smass2.html$\rangle$.}
(SMASSII; Bus and Binzel 2002) and the Small Solar System Objects Spectroscopic
Survey\footnote{Lazzaro, D., Angeli, C. A., Carvano, J. M., Mothe-Diniz, T., Duffard, R., and
Florczak, M., Small Solar System Objects Spectroscopic Survey V1.0. EAR-A-I0052-8-S3OS2-V1.0. NASA
Planetary Data System, 2006. $\langle$http://sbn.psi.edu/pds/resource/s3os2.html$\rangle$.}
(S$^3$OS$^2$; Lazarro et al. 2004).
The S-like asteroids include the S-complex and the minor classes including Ld-, D-, T-, V-, and
O-types.
The C-like asteroids consist of the C- and X-complexes.
The mean value is $p_{\rm{s}}$ = 0.19 $\pm$ 0.09 in the S-like asteroids (353 objects) and
$p_{\rm{c}}$ = 0.08 $\pm$ 0.07 in the C-like asteroids (649 objects).

The population ratio of S-like to C-like asteroids varies with heliocentric distance.
Yoshida and Nakamura (2007) showed that the ratio for asteroids of $D >$ 0.6 km are 1:1 in the
inner belt (2.0--2.6 AU), 1:2.3 in the middle belt (2.6--3.0 AU), and 1:4 in the outer belt
(3.0--3.5 AU).
We estimated the mean albedo in each of the three zones from the population ratio and mean albedos
of the S-like and C-like MBAs.
We adopted the mean albedos as
($p_{\rm{s}}$+$p_{\rm{c}}$)/(1+1) = 0.14 for 2.0 $< a <$ 2.6 AU,
($p_{\rm{s}}$+2.3$p_{\rm{c}}$)/(1+2.3) = 0.11 for 2.6 $< a <$ 3.0 AU, and
($p_{\rm{s}}$+4$p_{\rm{c}}$)/(1+4) = 0.10 for 3.0 $< a <$ 3.3 AU.
Although, as discussed in Section 5.2.2, the population of S-like MBAs is small in high
inclination, the difference is negligible with respect to mean albedos.
The absolute magnitude is converted to diameter using these albedo values and Eq.(4).

Figure 5 shows the diameter distribution for the detected MBAs.
Most of the bodies were smaller than 1 km in diameter.
Considering the photometric error of $\sim$0.1 mag and semi-major axis error of $\sim$0.1 AU,
the inaccuracy of diameter estimation is about 15\%.
When the error on the albedo ($\sim$0.06) is included, the added inaccuracy is 15\%.

\section{Discussion}

\subsection{Size Distributions}

\subsubsection{Size distribution of small MBAs}

The detected MBA sample included some observational bias, which had to be removed.
The most remarkable bias was a decrease in detection probabilities for faint objects.
The limiting magnitude (24.0 mag in the $r$-band or 23.8 mag in the $R_c$-band) corresponded
to $D$ = 0.7 km at a semi-major axis of 3.3 AU with a typical phase angle of 25$^{\circ}$
and a typical albedo of 0.1.

We regarded asteroids with $D >$ 0.7 km as the sample for estimating size distribution.
In addition, we excluded asteroids with $a <$ 2.5 AU for two reasons.
First, this zone contains only a small population of high-$i$ MBAs and S-like asteroids are
abundant in this region, unlike in the outer zone.
If these asteroids were included in the sample, the fraction of S-like asteroids would differ
significantly from the low-$i$ to high-$i$ MBAs.
S-like and C-like asteroid size distributions are supposed to differ
(Ivezi\'c et al. 2001, Yoshida \& Nakamura 2007, Wiegert et al. 2007).
An excess of S-like population in the low-$i$ MBAs could cause a bias in comparing size
distributions between low-$i$ and high-$i$ MBAs.
This bias is discussed again in Section 5.2.2.
Second, the estimated semi-major axis for an asteroid with $a <$ 2.5 AU has a large error
($>$0.15 AU).
Ultimately, 178 asteroids with $D >$ 0.7 km were selected in $a$ = 2.5--3.3 AU. 
The relationship between the sub-km asteroid population and inclination was examined with this
unbiased sample.

Figure 6 shows the fractions of the sub-km asteroids for each inclination bin of 5$^{\circ}$.
The fractions are almost constant ($\sim$0.4) between $0^{\circ}$ and $15^{\circ}$, and
significantly decrease beyond 15$^{\circ}$.
We found a population deficiency of sub-km MBAs with $i > 15^{\circ}$.

Figure 7 shows the cumulative size distributions (CSDs) for the low-$i$ sample with
$i < 15^{\circ}$ and high-$i$ sample with $i > 15^{\circ}$.
Bin size is $\Delta \log D [\rm{km}]$ = 0.05, indicating the uncertainty of the diameter
estimation.
The CSD is described by the power-law equation as
\begin{equation}
N(>D) \propto D^{-b},
\end{equation}
where $N(>D)$ is the cumulative number of asteroids larger than a diameter $D$ and $b$ is the
power-law index of the CSD. 
The single power law was fit to each CSD with 0.7 km $<D<$ 2 km by the least squares
method.
The best-fit indexes of the CSDs were $b$ = 1.79 $\pm$ 0.05 for the low-$i$ MBAs and
$b$ = 1.62 $\pm$ 0.07 for the high-$i$ MBAs.

We performed statistical tests to compare the two regressions.
The regression lines are described by $y = y_0 - bx$, where $x = \log D$, $y = \log N(>D)$,
and $y_0 = \log N(>1 \rm{km})$.
First, a $F$-test was used to check whether the two populations possessed homogeneous variances
around the regression lines.
We defined the null hypothesis as H$_0$: $\sigma_{\rm 1}^2 = \sigma_{\rm 2}^2$, against
the alternate hypothesis as H$_1$: $\sigma_{\rm 1}^2 \neq \sigma_{\rm 2}^2$, where $\sigma^2$ is
a residual variance and the suffix 1 and 2 stand for the low-$i$ and high-$i$ MBAs,
respectively.
The computed $F$-value was $\sigma_{\rm 2}^2/\sigma_{\rm 1}^2 =$ 2.00
($\sigma_{\rm 1}^2<\sigma_{\rm 2}^2$) less than the critical value $F_{0.05}$=3.18.
The test accepted that $\sigma_{\rm 1}^2$ and $\sigma_{\rm 2}^2$ were equal with the 0.05
level of significance.

Next, we performed a $t$-test with H$_0$: $b_1$=$b_2$ against H$_1$: $b_1 > b_2$.
Let the population regression coefficient for $b$ be normally distributed with mean $b_i$ and
variance $V_i\sigma_i^2$ where $V_i = 1/\Sigma (x-\overline{x}_i)^2$ and
$\overline{x}_i$ is the mean of $x_i$ ($i$=1,2).
In the case of equal variances, we use the criterion
\begin{equation}
t = \frac{b_1 - b_2}{\sqrt{\frac{(\Sigma_1+\Sigma_2)(V_1+V_2)}{f_1+f_2}}},
\end{equation}
where $\Sigma_1$ and $\Sigma_2$ are sums of squares from the regression lines, derived from 
$\Sigma_i = \Sigma [y - (y_0 - b_ix)]^2$ ($i$=1,2), and $f_1$ and $f_2$ are degrees of freedom
\citep{W38}.
The $t$-value for this test was calculated to be $t$=1.99 greater than the critical value
$t_{0.05}$=1.76 with 14 degree of freedom.
We rejected the null hypothesis and accepted $b_1 > b_2$ at the 0.05 level of significance.
In conclusion, for asteroids of at least 1 km in diameter, the high-$i$ MBAs had a shallower CSD
than the low-$i$ MBAs at the 95\% confidence level.

\subsubsection{Size distribution of large MBAs}

We used the published asteroid catalogs to obtain the size distribution of MBAs larger than those
detected in the Subaru data.
The CSDs for $D >$ 10 km were constructed using the Asteroid Orbital Elements Database
(ASTORB)\footnote{$\langle$http://www.naic.edu/$^\sim$nolan/astorb.html$\rangle$.}
distributed by Lowell Observatory \citep{B01}.
In addition, the SDSS Moving Object
Catalog\footnote{$\langle$http://www.astro.washington.edu/users/ivezic/sdssmoc/sdssmoc.html$\rangle$.}
(SDSS MOC; Ivezi\'c et al. 2002) was used for diameter in the 2--10 km range to link the CSDs
based on the Subaru and ASTORB data.

The ASTORB lists the orbital parameters and absolute magnitudes of more than 457,000 MBAs.
We regarded 453,439 asteroids as MBAs for which the semi-major axis was between 2.1 AU and 3.3 AU,
the perihelion was larger than 1.66 AU, and the aphelion was smaller than 4.61 AU \citep{Bo02}.
The diameter and albedo of 2228 MBAs were measured by IRAS observations in the IMPS/SIMPS surveys
\citep{T02}.
They cover almost all MBAs with $D >$ 35 km.
For MBAs of unknown albedo, the mean albedos deduced in Section 4.4 were used to estimate
diameters.
Figure 8(a) shows the absolute magnitude distribution and CSD of the MBAs in the ASTORB.
Although the detection limit was never determined because the asteroids were found by several
surveys with various limiting magnitudes, it appears to be a little under 10 km in diameter
as seen in the absolute magnitude distribution.

For km-size asteroids, the SDSS MOC was employed.
This lists the astrometric and photometric data for moving objects brighter than 21.5 mag in the
$r$-band.
The fourth released version of SDSS MOC consists of 470,000 moving objects observed up to March
2007.
The orbital elements of most SDSS MOC asteroids have not been determined.
The semi-major axis and inclination of each moving object in the SDSS MOC were estimated from
apparent motions assuming a circular orbit, as in the analysis of the Subaru data
(see Section 4.3).
For estimation accuracy, we used only objects with a phase angle smaller than 15$^{\circ}$.
A total of 130,000 objects located in 2.0--3.3 AU in the semi-major axis were defined as MBAs.

The diameter and albedo of each SDSS MOC asteroid are also unknown.
The apparent magnitudes of the moving objects were measured almost simultaneously in five bands
($u$, $g$, $r$, $i$, and $z$).
Ivezi\'c et al. (2001) defined a color index $a^{\ast}$ using $g - r$ and $r - i$ colors as
\begin{equation}
a^{\ast} \equiv 0.89(g - r) + 0.45(r - i) - 0.57
\end{equation}
to divide asteroids into two major groups, ``red" asteroids with $a^{\ast} > 0$ and ``blue"
asteroids with $a^{\ast} < 0$.
We considered the ``red" and ``blue" asteroids as S-like and C-like asteroids, respectively.
The diameter was obtained in the manner discussed in Section 4.4.
The albedo was given as 0.19 for S-like asteroids and 0.08 for C-like asteroids.
Figure 8(b) shows the absolute magnitude distribution and CSD in the SDSS MOC.
According to the absolute magnitude distribution, the detection limit was $H_r \sim 16.5$ mag,
which corresponds to $D \sim 2$ km at the outer edge of the main belt with the albedo of C-like
asteroids.

\subsubsection{Combined CSDs}

Finally, we constructed the continuous CSDs using the ASTORB for $D > 10$ km, the SDSS MOC for
2 km $< D <$ 10 km, and Subaru data for 0.7 km $< D <$ 2 km.
We used the only asteroids with $a$ = 2.5--3.3 AU to set the same sample condition as the analysis
of the Subaru data in Section 5.1.1.
The CSDs from the SDSS MOC were scaled to the cumulative number of the ASTORB sample at
$D$ = 10 km.
The CSD curves derived from the ASTORB and SDSS MOC connected smoothly in both the low-$i$ and
high-$i$ samples.
Likewise, the CSDs from the Subaru data were scaled to the cumulative number of the SDSS MOC
sample at $D$ = 2 km.

Figure 9 shows the combined CSDs of low-$i$ ($i < 15^{\circ}$) and high-$i$ ($i > 15^{\circ}$)
MBAs.
The CSD slope of each sample from 0.7 km to 50 km in diameter was $b$ = 2.17 $\pm$ 0.02 for
low-$i$ MBAs and $b$ = 2.02 $\pm$ 0.03 for high-$i$ MBAs.
The high-$i$ MBAs had a shallower CSD than the low-$i$ MBAs, at least down to $D$ = 0.7 km.
The difference in $b$ was not caused by the wavy patterns seen in Figure 9.
For confirmation we adjusted the high-$i$ CSD to the low-$i$ CSD at $D$ = 50 km, and its power-law
index was transformed to the same value as the low-$i$ CSD (see Fig 10).
The adjusted CSD of high-$i$ MBAs agreed well with the low-$i$ CSD over all sizes.
To state the significance of this result, we performed a $t$-test (see Section 5.1.1) with the null
hypothesis H$_0$: a regression line of the CSD difference, $y=y_0-bx$ where
$y = \log N_1(>D) - \log N_2(>D)$, is not significant ($b = 0$) and the alternative hypothesis
$H_1$: the regression line is significant ($b > 0$).
The $t$-value is $b/\sqrt{\Sigma V f^{-1}} =$ 16.2 much greater than $t_{0.01}$ (2.43), which
rejected the null hypothesis at the 99\% confidence level.
This test indicated that the overall high-$i$ CSD was definitely shallower than the low-$i$ CSD.

\subsection{Differences in the CSD Slopes}

\subsubsection{Collision velocities}

Collisional fragmentation is characterized by impact strength, $Q^{\ast}_{\rm D}$.
$Q^{\ast}_{\rm D}$ is the critical specific energy to fragment the target and disperse half of its
mass.
It depends on asteroid size, generally expressed as
\begin{equation}
Q^{\ast}_{\rm D} \propto D^{s}.
\end{equation}
The $Q^{\ast}_{\rm D}$ curve has two different power laws with opposite signs.
For small bodies, $Q^{\ast}_{\rm D}$ decreases with increasing diameter ($s < 0$), in what is
called the strength-scaled regime.
In contrast, strength increases with increasing diameter ($s > 0$) for large bodies, in what is
called the gravity-scaled regime.
The two slopes join at the transition diameter ($D_t$) of 0.1--1 km \citep{D98}.
O'Brien and Greenberg (2003) showed that the power-law index of the CSD with $D > D_t$ is
determined by the power-law index of the $Q^{\ast}_{\rm D}$ curve in the gravity-scaled
regime, $s_{\rm g}$.
According to the collisional steady-state scenario, this relationship is described by
\begin{equation}
b = \frac{5}{2+s_{\rm g}/3}.
\end{equation}

In the present study, the overall CSD slope of high-$i$ MBAs ($b$ = 2.02) was smaller than that of
low-$i$ MBAs ($b$ = 2.17), corresponding to $s_{\rm g}$ = 0.91 and $s_{\rm g}$ = 1.43,
respectively.
The most remarkable difference between the low-$i$ and high-$i$ MBAs was collision velocity.
The mean relative velocity among MBAs, most of which are low-$i$ bodies, is 3.8 km s$^{-1}$
\citep{V98}.
In contrast, high-$i$ MBAs often collide at high velocities, around 10 km s$^{-1}$ \citep{G06}.
We propose that these hypervelocity collisions have a large $s_{\rm g}$.

However, such $s_{\rm g}$ has not been determined through previous impact experiments nor
through numerical simulations.
Benz and Asphaug (1999) performed impact simulations for basalt at two different collision
velocities, 3 km s$^{-1}$ and 5 km s$^{-1}$, and found that $s_{\rm g}$ did not vary between the
two velocities.
However, Jutzi et al. (2010), using impact simulations, reported that the $s_{\rm g}$ of basalt
targets for 5 km s$^{-1}$ ($s_{\rm g}$ = 1.29) was smaller than that for 3 km s$^{-1}$
($s_{\rm g}$ = 1.33).
These values correspond to $b$ = 2.06 and $b$ = 2.05, respectively.
This difference is too small to be recognized through observations.
Their results showed that $s_{\rm g}$ does not depend strongly on collisional velocity enough to
shift $b$.
But the relationship between the $Q^{\ast}_{\rm D}$ curve and collision velocity, especially in
the hypervelocity collisions around 10 km s$^{-1}$, is still uncertain.
Our study might present evidence for an increase in $s_{\rm g}$ in hypervelocity
collisions around 10 km s$^{-1}$.

\subsubsection{Taxonomic distribution}

Examined qualitatively, it is possible that the different spatial distributions between taxonomic
groups caused the discrepancy in the CSD slopes between low-$i$ and high-$i$ MBAs.
The population ratio of S-like to C-like asteroids is not uniform with inclination.
Moth\'e-Diniz et al. (2003) showed that a smaller fraction of S-complex MBAs with $D >$ 7 km
appears in the high eccentricity/inclination regime.
Meanwhile, several observations have indicated different CSD slopes for different taxonomic groups,
mostly in low inclination (Ivezi\'c et al. 2001, Yoshida \& Nakamura 2007, Wiegert et al. 2007).
This difference in abundance between the taxonomic groups could induce an apparent decrease in the
CSD slope in high inclination.

We checked this effect using the SDSS MOC data.
The SDSS asteroids were separated into S-like and C-like groups by $a^\ast$ color defined
in Eq.(6).
The low-$i$ MBAs included 6585 S-like asteroids and 9746 C-like asteroids.
The high-$i$ MBAs include 1139 S-like asteroids and 2019 C-like asteroids.
The CSDs of the low-$i$ and high-$i$ MBAs (2.5 AU $< a <$ 3.3 AU) were constructed in each group.
The power-laws for these CSDs were approximated for the 2--10 km in diameter range.
Among the S-like asteroids, the slope, named $b_{\rm s}$, was 2.66 $\pm$ 0.04 for low-$i$ MBAs
and 2.26 $\pm$ 0.03 for high-$i$ MBAs.
Among the C-like asteroids, the slope, named $b_{\rm c}$, was 2.41 $\pm$ 0.02 for low-$i$ MBAs
and 2.36 $\pm$ 0.05 for high-$i$ MBAs.
The $b_{\rm s}$ value was significantly smaller in high inclination, while $b_{\rm c}$ was similar
for low-$i$ and high-$i$ MBAs.

We evaluated the contribution of the S-like MBAs to the decrease in $b$ of high-$i$ MBAs.
Model CSDs were compounded from artificial power-law CSDs of S-like and C-like MBAs based on a
population model.
The model was characterized by a population ratio of S-like MBAs to C-like MBAs and their CSD
slopes (Table 3).
We compared indexes of fitted power laws to the model CSDs with those of the observed CSDs using
two models.
In model 1, we explored the possibility that the small $b_{\rm s}$ of high-$i$ S-like MBAs was a
major factor in the small $b$ of total high-$i$ MBAs.
The population ratio of S-like asteroids was set to the same value as the SDSS MOC data, 0.67
(6585/9746) for low-$i$ MBAs and 0.56 (1139/2019) for high-$i$ MBAs.
The $b_{\rm s}$ and $b_{\rm c}$ values for high-$i$ MBAs were set to match the low-$i$ MBAs
($b_{\rm s}$=2.66, $b_{\rm c}$=2.41).
The model CSDs showed $b =$ 2.50 for low-$i$ MBAs and $b =$ 2.49 for high-$i$ MBAs.
The CSD slope of high-$i$ MBAs was not as small as that of the observed CSD.
Consequently, these results show that a small (compound) slope $b$ in the high-$i$ MBAs requires a
smaller S-type slope $b_{\rm s}$ for that region.
In model 2, we examined the possibility that the small S-like/C-like population ratio of high-$i$
MBAs caused the small $b$.
The test was performed assuming that high-$i$ MBAs have the same population ratio as low-$i$ MBAs.
The CSD power-law indexes were set to match the SDSS MOC data, ($b_{\rm s}$, $b_{\rm c}$) =
(2.66, 2.41) for low-$i$ MBAs and ($b_{\rm s}$, $b_{\rm c}$) = (2.26, 2.36) for high-$i$ MBAs.
The best-fit power-law indexes of the compounded CSDs were $b =$ 2.50 for low-$i$ MBAs and
$b =$ 2.32 for high-$i$ MBAs.
The both of $b$ values agreed with the observed CSDs.
This indicates that the scarcity of S-like asteroids with high inclination had little influence on
the CSD slope.
We conclude that the shallow CSD of the high-$i$ MBAs was due to the small CSD slope of S-like MBAs
with high inclination.

We suggest that $s_{\rm g}$ values for high-velocity collisions (around 10 km s$^{-1}$) of S-like
asteroids are large, while $s_{\rm g}$ values for similar C-like asteroids are not.
One possible cause is the variation in porosities between S-like and C-like asteroids.
C-type asteroids have porosities of 35--60\%, whereas S-type asteroids have porosities of around
30\% \citep{Br02}.
Jutzi et al. (2010) showed through impact simulations that $s_{\rm g}$ is constant (1.22)
between collision velocities of 3 km s$^{-1}$ and 5 km s$^{-1}$ for porous targets (pumice).
On the other hand, the $s_{\rm g}$ values for 5 km s$^{-1}$ collisions (1.29) is smaller than
3 km s$^{-1}$ collisions (1.36) for the non-porous targets (basalt).
Porosity may have a resistance effect to $s_{\rm g}$ variation with collision velocity.

Note that these considerations include uncertainty.
The wavy structure of the CSD may differ between S-like and C-like MBAs.
Estimating the CSD slopes from a narrow size range is not appropriate for comparing the overall
slope $b$.
Indeed, previous surveys have presented a variety of results.
Ivezi\'c et al. (2001) showed $b$ = 3.00 $\pm$ 0.05 for both groups with $D \geqq$ 5 km, and
$b$ = 1.20 $\pm$ 0.05 for S-like asteroids and $b$ = 1.40 $\pm$ 0.05 for C-like asteroids
with $D \leqq$ 5 km.
Yoshida and Nakamura (2007) showed $b$ = 2.44 $\pm$ 0.09 for S-like asteroids with 1 km $<D<$ 3 km
and $b$ = 1.29 $\pm$ 0.02 with 0.3 km $<D<$ 1 km, and a single slope $b$ = 1.33 $\pm$ 0.03 for
C-like asteroids with 0.6 km $<D<$ 7 km.
The difference in CSD between the S-like and C-like asteroids is still unclear.

\subsubsection{Dynamical removal}

Dynamical processes, as well as collisional processes, affect the shape of the CSD.
O'Brien and Greenberg (2005) showed via numerical simulations that the CSD slope of MBAs is
reduced from $\sim$20 km in diameter through the action of the Yarkovsky effect and
resonances.
The Yarkovsky effect causes a drift in the semi-major axis, the rate of change of which increases
with decreasing size down to 0.01 km in diameter.
Asteroids caught in a strong resonance are removed from the main belt.
Smaller asteroids are removed more quickly from the main belt.
Although the Yarkovsky effect is independent of inclination, several secular resonances appear at
high inclination.
Within the heliocentric distance of 2.5--3.3 AU, none of these resonances appear at
$i < 15^{\circ}$.
In contrast, the strong secular resonances of $\nu_5$, $\nu_6$, and $\nu_{16}$ lie in
$i > 15^{\circ}$ \citep{K91}.
High-$i$ asteroids in this zone are subject to removal.
These secular resonances together with the Yarkovsky effect could cause significant losses of high-$i$
MBAs, possibly enough to diminish the CSD slope.
This action increases the difference in $b$ between the low-$i$ and high-$i$ MBAs at small sizes
($D \leqq 20$ km).
However, as shown in Figure 10, the difference in $b$ is almost constant as the diameter decreases
from $D \sim 50$ km.
This is inconsistent with the above expectation.
Hence, dynamical removal does not decrease $b$ in high inclination.

\section{Conclusion}

We performed wide-field observations of high-$i$ MBAs of sub-km diameter using the 8.2 m Subaru
Telescope.
The CSDs of low-$i$ and high-$i$ MBAs were constructed using the Subaru data and the
ASTORB and SDSS MOC catalogs.
A summary of our main conclusions follows.

(1) A smaller fraction of small asteroids (0.7 km $<D<$ 1 km) appear in high inclination.
The power-law index of the CSD for the 0.7--2 km diameter range is 1.79 $\pm$ 0.05 for low-$i$ MBAs
($i < 15^\circ$) and 1.62 $\pm$ 0.07 for high-$i$ ($i > 15^\circ$) MBAs.
The sub-km asteroids at $i > 15^\circ$ have a lower index than the low-$i$ sample at the
95\% confidence level based in a statistical $t$-test.

(2) The single power-law slope of the combined CSD for diameter in the 0.7--50 km range is 2.17
$\pm$ 0.02 for the low-$i$ MBAs and 2.02 $\pm$ 0.03 for the high-$i$ MBAs.
The high-$i$ CSD is shallower than the low-$i$ CSD for the entire size range at the 99\%
confidence level.
This is not caused by the difference in the wavy pattern of the CSDs.

(3) The CSD of S-like asteroids has a small slope in high inclination, whereas that of C-like
asteroids shows little variation in slope with inclination.
Through modeling we showed that the shallow CSD of the high-$i$ MBAs is not caused by the spatial
distribution of the taxonomic groups, but by the shallow CSD of the S-like MBAs with high
inclination.

(4) The difference in slope of CSDs between the low-$i$ and high-$i$ MBAs is constant across
$D \sim 10$ km.
The shallow CSD of the high-$i$ MBAs is not the result of dynamical removal due to the Yarkovsky
effect and the secular resonances.

We suggest that the small $b$ of S-like MBAs with high inclination is due to a collisional effect.
The possible explanation is that the $Q^{\ast}_{\rm D}$ curve has a large gravity-scaled regime
$s_{\rm g}$ slope under hypervelocity collisions (around 10 km s$^{-1}$).
Asteroid collisions often occurred with such high velocities in the dynamical excitation phase in
the final stage of planet formation.
We suppose that during this phase, MBAs experienced oligopolistic collisional evolution; small
bodies were more easily disrupted relative to large bodies than at present.
This indication claims that the current evolutionary models for MBAs should be modified.

\bigskip

\section*{Acknowledgments}

This work was partly supported by ``The 21st Century COE Program: The Origin and Evolution of
Planetary Systems" of the MEXT.
T. Terai was supported by the Grant-in-Aid from Japan Society for the Promotion of Science
(20-4879).

\clearpage

\begin{figure}
\begin{center}
\FigureFile(150mm,150mm){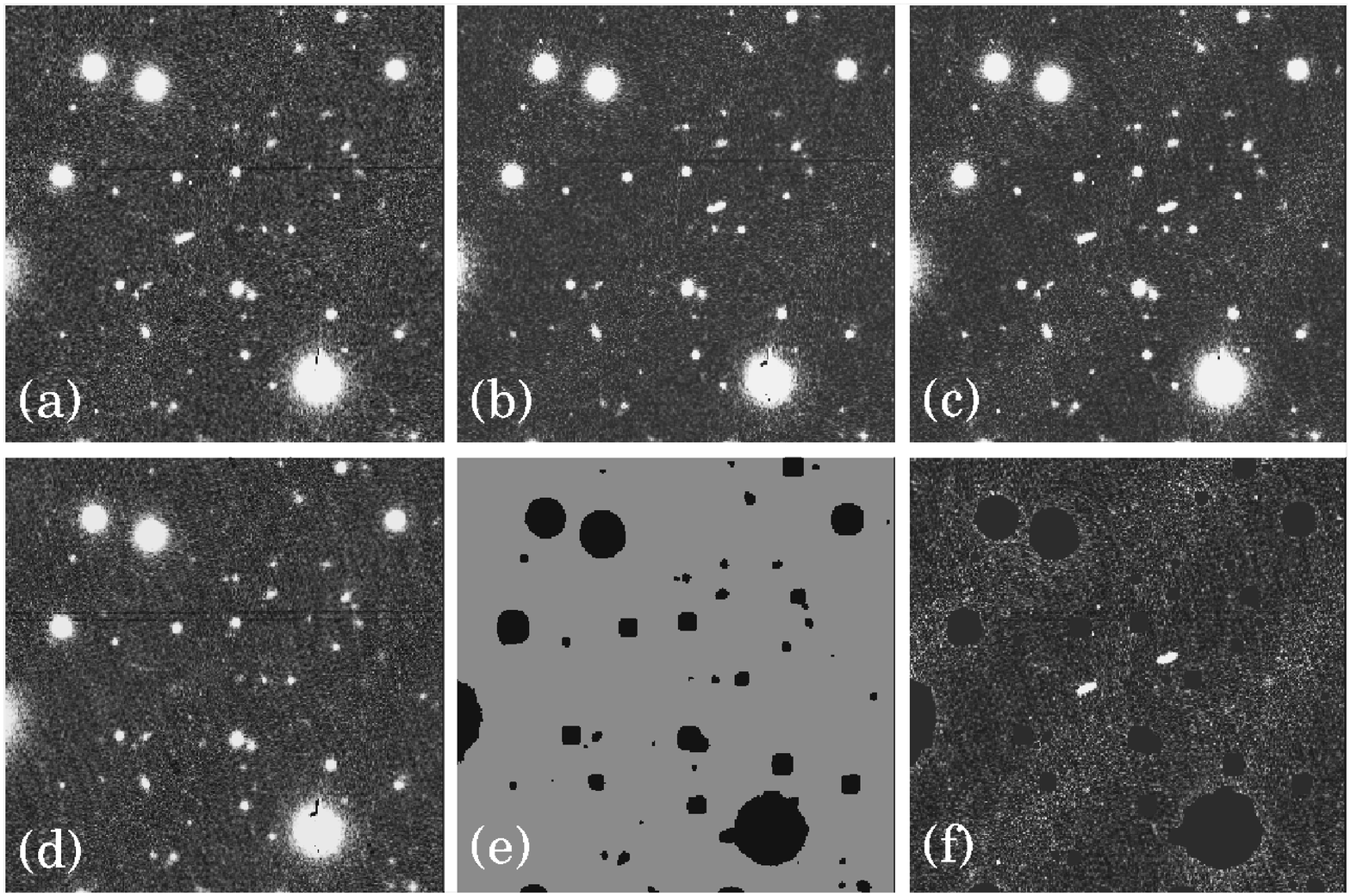}
\end{center}
\caption{
 The procedures of data reduction for detecting moving objects (see text). (a) and (b) are a
 portion of the original $r$-band images taken at a 20-minute interval.
 (c) and (d) are the ``OR" and ``AND" images, respectively.
 (e) shows the mask image created from the AND image.
 (f) is the final processed image.
 The field stars and galaxies are eliminated and the moving object appears as two sources.
 The field-of-view of each image is $1^{\prime} \times 1^{\prime}$.
 \label{fig1}}
\end{figure}

\clearpage

\begin{figure}
\begin{center}
\FigureFile(120mm,120mm){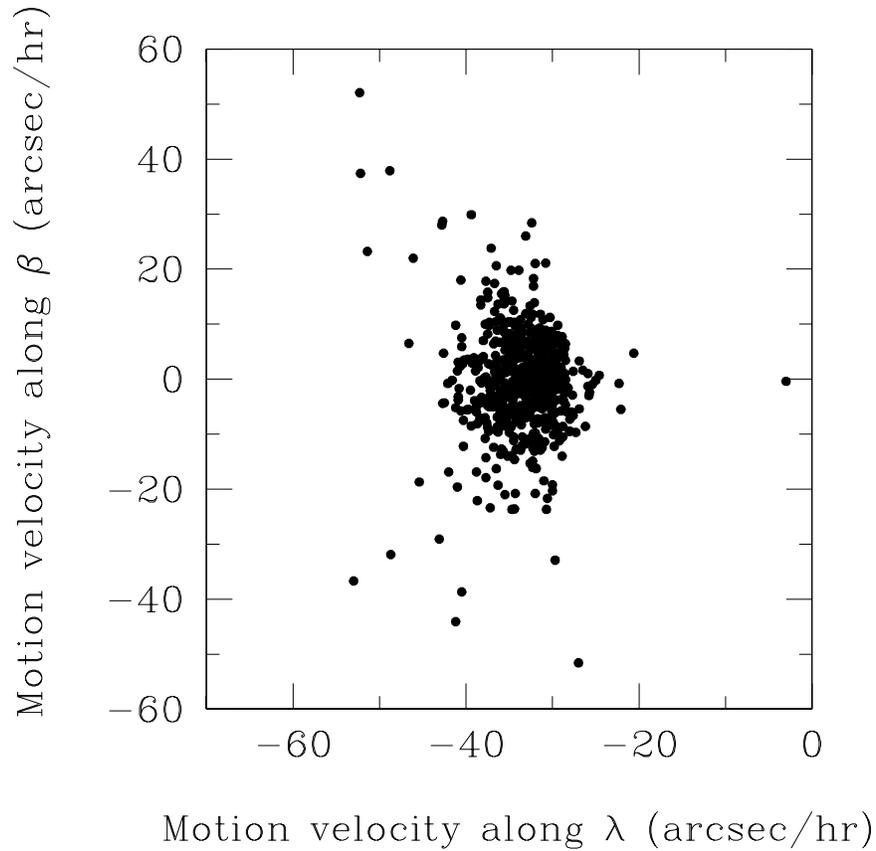}
\end{center}
\caption{
 Ecliptic longitude ($\lambda$) and latitude ($\beta$) components of the apparent motions of the
 detected moving objects.
 The largest swarm consists of MBAs with low inclination. 
 High-inclination MBAs are located away from 0 arcsec hr$^{-1}$ of the $\beta$ component.
 The point with low velocities corresponds to a trans-Neptunian object.
 \label{fig2}}
\end{figure}

\clearpage

\begin{figure}
\begin{center}
\FigureFile(150mm,150mm){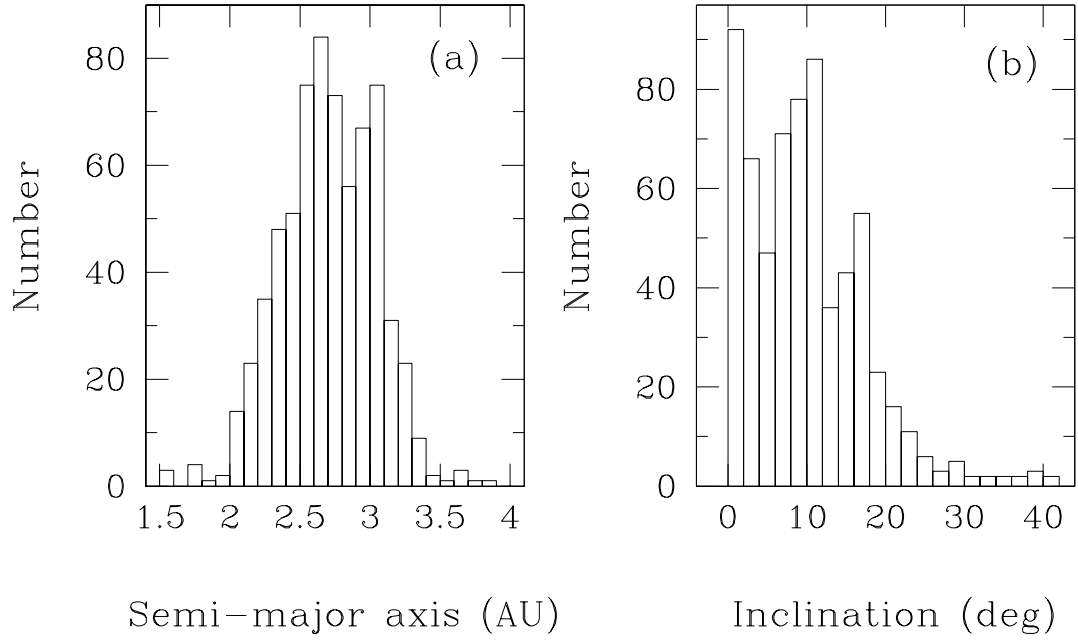}
\end{center}
\caption{
 The distributions in semi-major axis (left) and inclination (right) of the detected moving
 objects.
 \label{fig3}}
\end{figure}

\clearpage

\begin{figure}
\begin{center}
\FigureFile(150mm,150mm){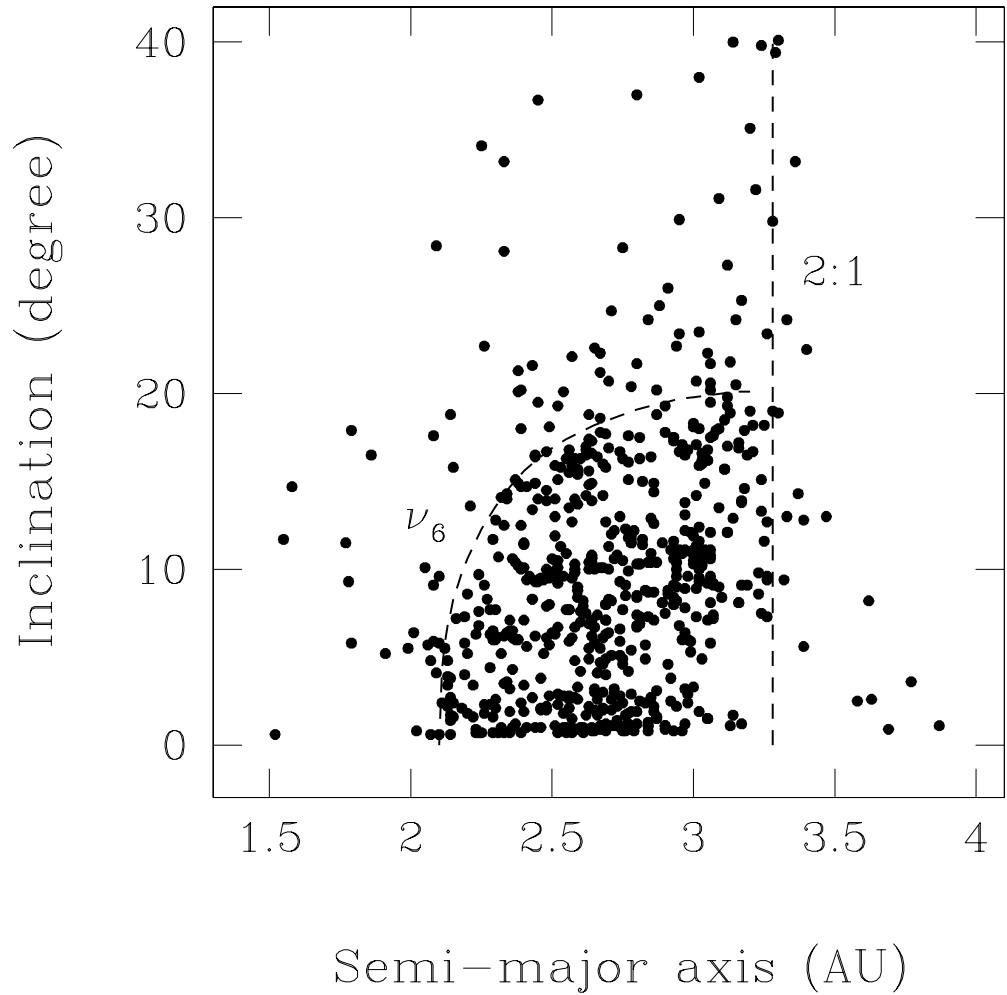}
\end{center}
\caption{
 Semi-major axis vs. inclination distribution of the detected moving objects.
 The dashed vertical line shows the 2:1 mean motion resonance, and the dashed curve shows the
 $\nu_6$ secular resonance.
 \label{fig4}}
\end{figure}

\clearpage

\begin{figure}
\begin{center}
\FigureFile(120mm,120mm){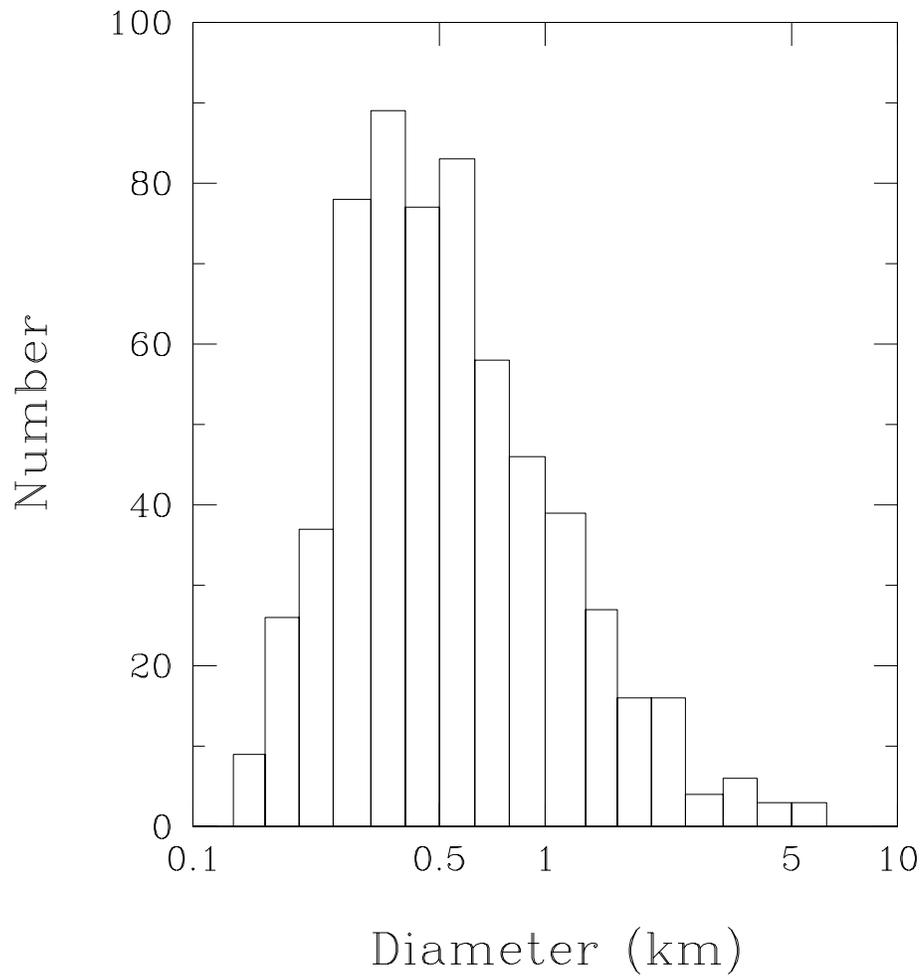}
\end{center}
\caption{
 Diameter distribution of the detected MBAs.
 The horizontal axis is shown in logarithmic scale with a step of 0.1.
 \label{fig5}}
\end{figure}

\clearpage

\begin{figure}
\begin{center}
\FigureFile(150mm,150mm){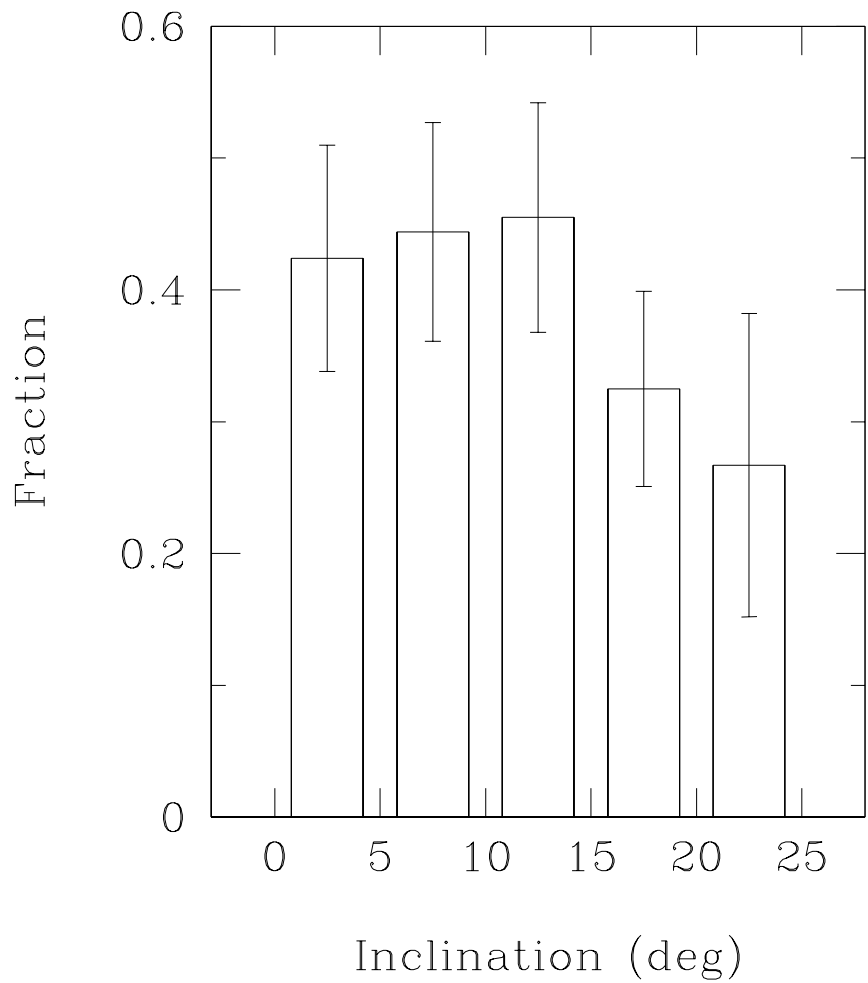}
\end{center}
\caption{
 The fraction of MBAs smaller than 1 km in diameter.
 This drops steeply beyond 15$^{\circ}$.
 \label{fig6}}
\end{figure}

\clearpage

\begin{figure}
\begin{center}
\FigureFile(150mm,150mm){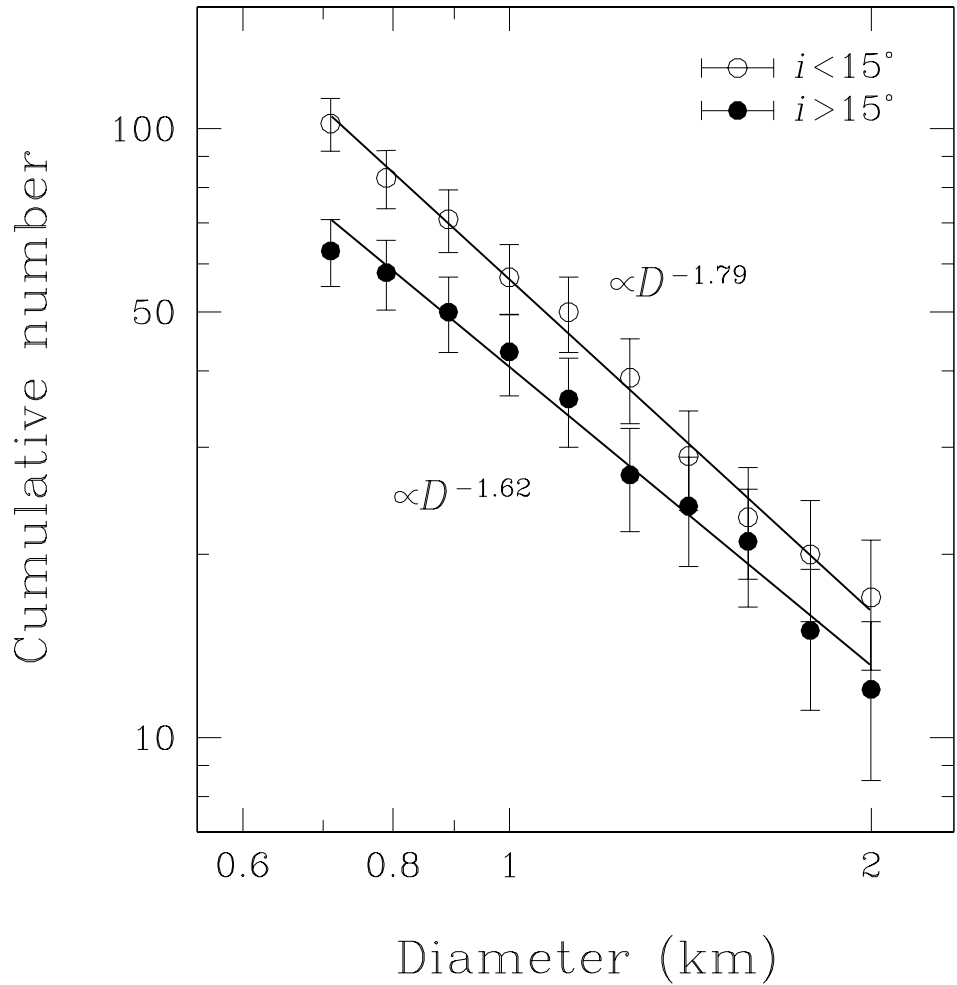}
\end{center}
\caption{
 Cumulative size distributions (CSDs) of the MBA sample.
 The open circles show the CSD of low-$i$ MBAs ($i<15^{\circ}$) and
 the filled circles show the CSD of high-$i$ MBAs ($i>15^{\circ}$).
 The solid lines are fitted power laws of $N(>D) \propto D^{-b}$.
 The slope of the high-$i$ CSD ($b$ = 1.62) is smaller than that of the low-$i$ CSD ($b$ = 1.79).
 \label{fig7}}
\end{figure}

\clearpage

\begin{figure}
\begin{center}
\FigureFile(150mm,150mm){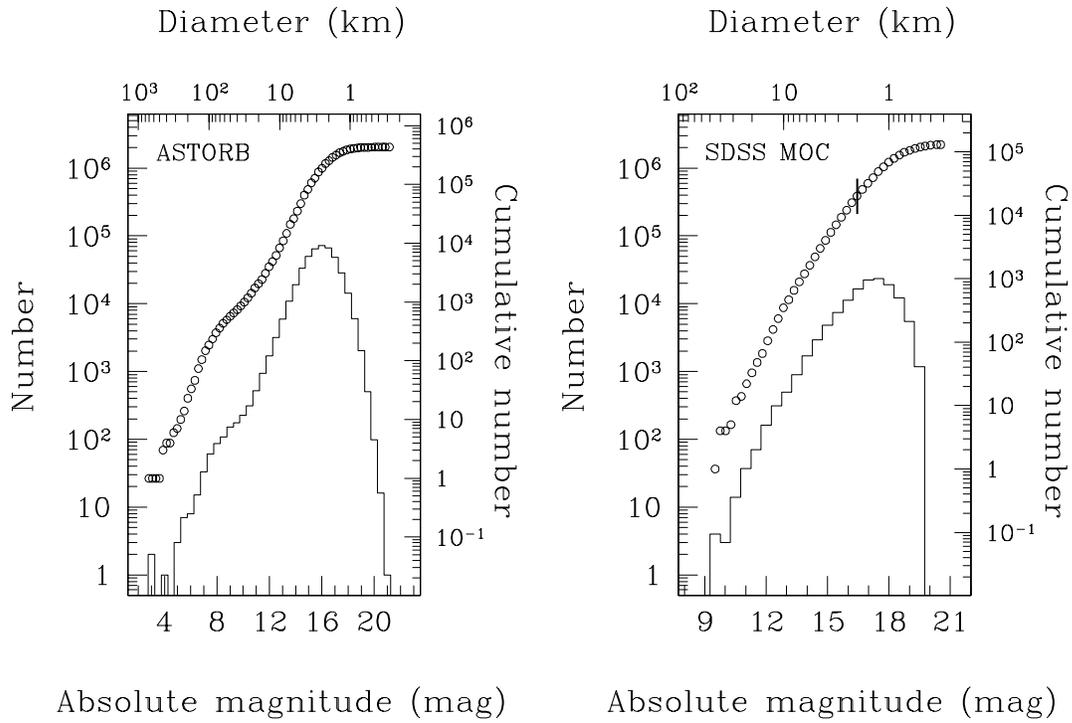}
\end{center}
\caption{
 The absolute magnitude distributions (histogram) and the cumulative size distributions
 (open circles) obtained by the ASTORB database in the left and SDSS Moving Objects Catalog
 (SDSS MOC) in the right.
 The horizontal axes of absolute magnitude (bottom) and diameter (upper) do not exactly
 correspond with each other.
 The vertical bar in the panel of the SDSS MOC shows a detection limit of 2 km in diameter.
 \label{fig8}}
\end{figure}

\clearpage

\begin{figure}
\begin{center}
\FigureFile(150mm,150mm){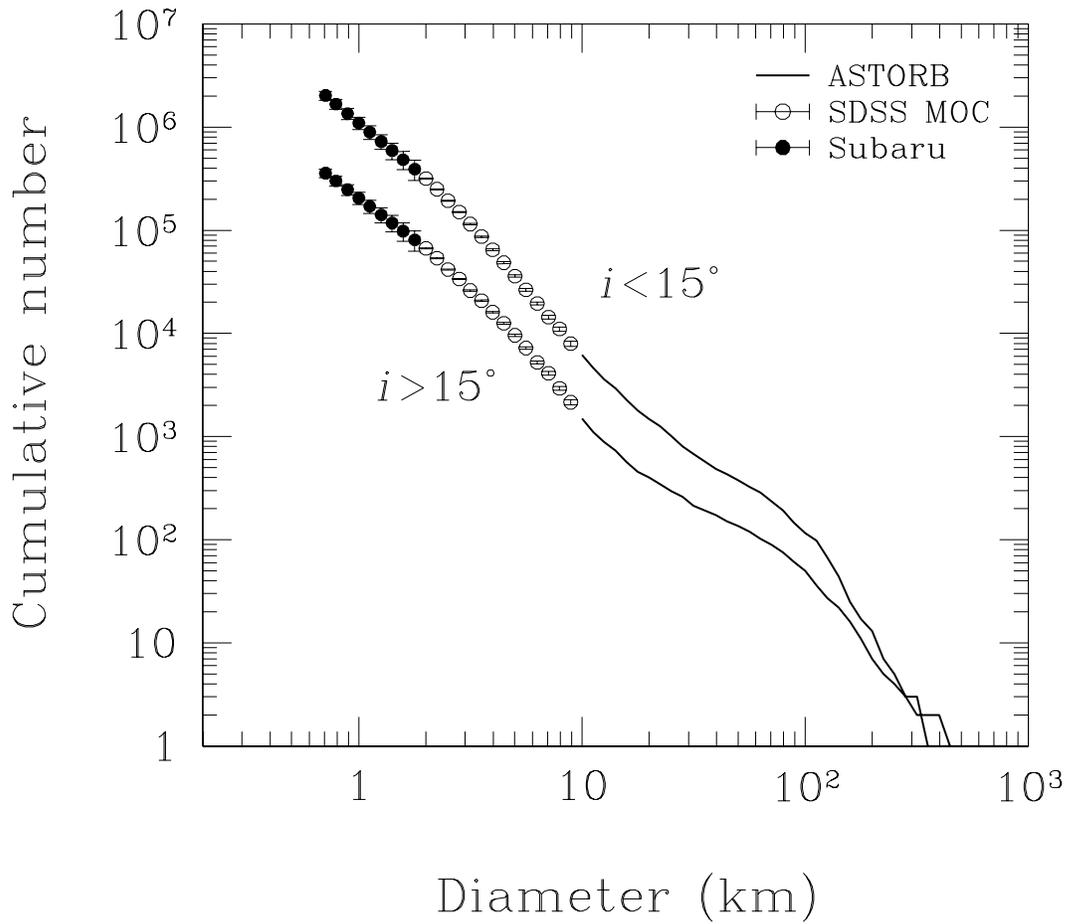}
\end{center}
\caption{
 Combined CSDs of the low-$i$ and high-$i$ MBAs down to 0.7 km in diameter.
 The solid curves are the numbered MBA population in the ASTORB database, the open circles show
 the scaled CSDs based on the SDSS MOC, and the filled circles show the scaled CSDs based on
 the Subaru data.
 The high-$i$ CSD is shallower than the low-$i$ CSDs across the entire size range.
 \label{fig9}}
\end{figure}

\clearpage

\begin{figure}
\begin{center}
\FigureFile(150mm,150mm){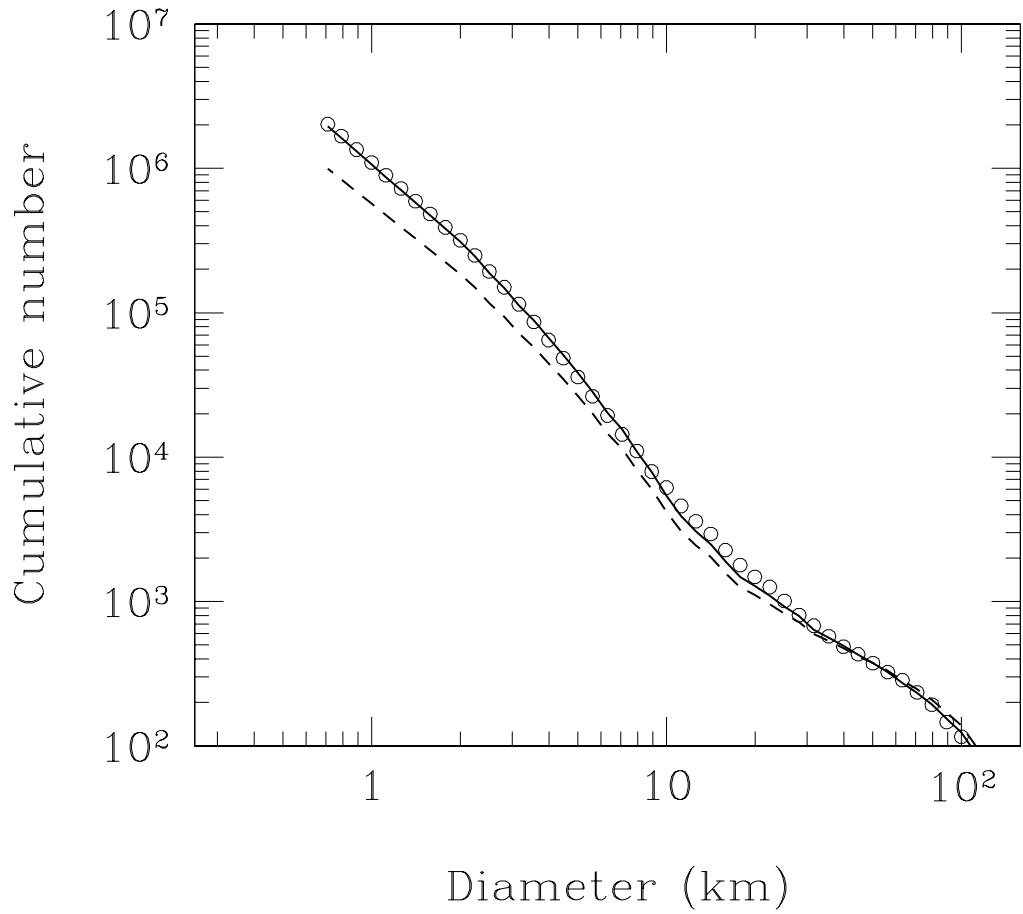}
\end{center}
\caption{
 Comparison of the CSDs between low-$i$ (open circles) and high-$i$ MBAs (dashed line).
 The solid curve is the adjusted high-$i$ CSD whose single-slope power-law index fit was
 transformed into the index of the low-$i$ CSD.
 It coincides with the low-$i$ CSD across the entire size range of less than 100 km in object
 diameter.
 \label{fig10}}
\end{figure}

\clearpage

\begin{table}
\begin{center}
\caption{Observations with the Suprime-Cam on June 3, 2008 (UT).\label{tbl-1}}
\begin{tabular}{ccccc}
\hline
 Field &
 RA(J2000) &
 Dec(J2000) &
 $\beta$\footnotemark[$*$] (deg) &
 Obs. time (UT) \\
\hline
Field 1-1 & 16:39:00 & +00:00:00 & +21.9 & 10:46:58, 11:12:50 \\
Field 1-2 & 16:36:00 & +00:36:00 & +22.4 & 10:52:02, 11:17:57 \\
Field 1-3 & 16:47:10 & +02:40:00 & +24.8 & 10:57:07, 11:22:59 \\
Field 1-4 & 16:49:25 & +01:25:00 & +23.7 & 11:02:40, 11:28:02 \\
Field 2-1 & 16:51:10 & +02:30:30 & +24.8 & 11:33:04, 11:53:11 \\
Field 2-2 & 16:53:00 & +02:03:33 & +24.4 & 11:38:07, 11:58:13 \\
Field 2-3 & 16:52:30 & +01:34:49 & +23.9 & 11:43:08, 12:03:17 \\
Field 2-4 & 16:52:35 & +01:04:59 & +23.5 & 11:48:08, 12:08:17 \\
Field 3-1 & 16:53:00 & +03:25:00 & +25.8 & 12:13:18, 12:33:28 \\
Field 3-2 & 16:55:30 & +03:15:00 & +25.7 & 12:18:22, 12:38:32 \\
Field 3-3 & 16:55:30 & +02:32:59 & +25.0 & 12:23:23, 12:43:34 \\
Field 3-4 & 16:55:30 & +01:55:30 & +24.4 & 12:28:27, 12:48:37 \\
Field 4-1 & 17:00:28 & +03:25:29 & +26.0 & 12:53:40, 13:13:47 \\
Field 5-1 & 17:02:30 & +01:58:00 & +24.6 & 13:34:00, 13:56:47 \\
\hline

\end{tabular}
\end{center}

\par\noindent
\footnotemark[$*$] Ecliptic latitude.

\end{table}

\clearpage

\begin{table}
\begin{center}
\caption{The Suprime-Cam archival data used in this survey.\label{tbl-2}}
\begin{tabular}{rrrrrrc}
\hline
 Field &
 RA(J2000) &
 Dec(J2000) &
 $\beta$\footnotemark[$*$] (deg) &
 Date (UT) &
 Obs. time (UT) &
 Exp. (sec) \\
\hline
SDA01 & 13:24:29  &   +27:46:13  &   +33.7 \ \ & 2002-04-11 \ &  06:49:10, 08:10:58 & 480 \\
SDA02 & 13:24:33  &   +27:29:22  &   +33.4 \ \ & 2003-04-02 \ &  09:57:21, 10:51:20 & 480 \\
SDA03 & 10:12:18  &   +46:56:40  &   +33.2 \ \ & 2004-02-17 \ &  08:03:01, 08:27:02 & 300 \\
SDA04 & 08:49:46  &   +44:15:14  &   +25.5 \ \ & 2002-01-09 \ &  14:10:52, 14:17:32 & 280 \\
SDA05 & 08:47:20  &   +44:15:15  &   +25.3 \ \ & 2002-01-09 \ &  14:24:05, 14:30:31 & 280 \\
SDA06 & 08:49:46  &   +44:15:15  &   +25.5 \ \ & 2002-01-15 \ &  14:05:25, 14:12:15 & 280 \\
SDA07 & 08:47:20  &   +44:15:15  &   +25.3 \ \ & 2002-01-15 \ &  14:19:01, 14:25:30 & 280 \\
SDA08 & 21:53:40  &   +17:39:52  &   +28.5 \ \ & 2004-09-15 \ &  06:21:46, 06:49:48 & 500 \\
SDA09 & 11:49:34  &   +22:23:42  &   +19.4 \ \ & 2005-03-05 \ &  08:37:07, 08:44:07 & 150 \\
SDA10 & 07:02:42  &   +38:54:40  &   +16.2 \ \ & 2002-01-12 \ &  06:47:56, 07:11:58 & 600 \\
SDA11 & 00:18:16  &   +16:17:40  &   +13.1 \ \ & 2003-09-27 \ &  08:23:14, 08:50:05 & 480 \\
SDA12 & 22:18:21  &   +00:37:27  &   +10.4 \ \ & 2004-08-18 \ &  09:13:12, 09:31:14 & 480 \\
SDA13 & 23:31:55  &   +00:10:38  &    +3.0 \ \ & 2003-09-28 \ &  05:26:50, 05:52:19 & 450 \\
SDA14 & 23:34:03  &   +00:10:38  &    +2.7 \ \ & 2003-09-28 \ &  06:03:10, 06:28:40 & 450 \\
SDA15 & 23:28:29  &   +00:08:39  &    +3.3 \ \ & 2003-09-28 \ &  06:38:23, 07:03:48 & 450 \\
SDA16 & 00:22:53  &   +04:25:14  &    +1.8 \ \ & 2003-09-28 \ &  07:26:54, 07:52:16 & 450 \\
SDA17 & 23:56:57  & $-$00:59:31  &  $-$0.6 \ \ & 2004-09-16 \ &  08:17:16, 08:35:19 & 300 \\
SDA18 & 23:01:27  & $-$05:34:30  &    +0.6 \ \ & 2004-09-16 \ &  08:53:01, 09:20:15 & 480 \\
SDA19 & 21:40:18  & $-$23:40:51  &  $-$9.2 \ \ & 2003-07-27 \ &  11:18:13, 11:36:07 & 480 \\
SDA20 & 07:50:50  &   +10:09:15  & $-$10.7 \ \ & 2002-01-09 \ &  13:43:14, 13:49:47 & 280 \\
SDA21 & 07:49:04  &   +10:09:15  & $-$10.7 \ \ & 2002-01-09 \ &  13:56:17, 14:02:55 & 280 \\
SDA22 & 07:50:49  &   +10:08:30  & $-$10.7 \ \ & 2002-01-14 \ &  13:39:28, 13:46:16 & 280 \\
SDA23 & 07:49:03  &   +10:08:30  & $-$10.7 \ \ & 2002-01-14 \ &  13:53:34, 14:00:05 & 280 \\
SDA24 & 02:18:03  & $-$04:58:29  & $-$17.7 \ \ & 2002-10-08 \ &  10:21:43, 10:39:37 & 480 \\
SDA25 & 02:18:00  & $-$05:25:00  & $-$18.1 \ \ & 2002-11-02 \ &  11:05:01, 11:22:55 & 480 \\
SDA26 & 02:18:00  & $-$04:35:00  & $-$17.3 \ \ & 2002-11-02 \ &  11:49:55, 12:07:50 & 480 \\
SDA27 & 02:19:47  & $-$05:00:00  & $-$17.9 \ \ & 2002-11-03 \ &  12:22:08, 12:40:03 & 480 \\
SDA28 & 02:16:21  & $-$04:59:00  & $-$17.6 \ \ & 2002-11-03 \ &  13:26:36, 13:53:41 & 480 \\
\hline

\end{tabular}
\end{center}

\par\noindent
\footnotemark[$*$] Ecliptic latitude.

\end{table}

\clearpage

\begin{table}
\begin{center}
\caption{The power-law indexes of the compounded CSDs from S-like and C-like MBAs. \label{tbl-3}}
\begin{tabular}{lcccc}
\hline
 &
 S/C\footnotemark[$*$] &
 $b_{\rm s}$\footnotemark[$\dagger$] &
 $b_{\rm c}$\footnotemark[$\ddagger$] &
 $b$\footnotemark[$\S$] \\
\hline
SDSS MOC\\
\ \ \ low-$i$  & 0.67 & 2.66 & 2.41 & 2.49$\pm$0.02 \\
\ \ \ high-$i$ & 0.56 & 2.26 & 2.36 & 2.32$\pm$0.04 \\
\\
Model 1\\
\ \ \ low-$i$  & 0.67 & 2.66 & 2.41 & 2.50\footnotemark[$\|$] \\
\ \ \ high-$i$ & 0.56 & 2.66 & 2.41 & 2.49\footnotemark[$\|$] \\
\\
Model 2\\
\ \ \ low-$i$  & 0.67 & 2.66 & 2.41 & 2.50\footnotemark[$\|$] \\
\ \ \ high-$i$ & 0.67 & 2.26 & 2.36 & 2.32\footnotemark[$\|$] \\
\hline

\end{tabular}
\end{center}

\par\noindent
\footnotemark[$*$] Population ratio of the S-like MBAs to the C-like MBAs.
\par\noindent
\footnotemark[$\dagger$] Power-law index of the CSD for S-like group.
\par\noindent
\footnotemark[$\ddagger$] Power-law index of the CSD for C-like group.
\par\noindent
\footnotemark[$\S$] Power-law index of the summed CSD.
\par\noindent
\footnotemark[$\|$] The fitting error is $\ll$ 0.01.

\end{table}

\end{document}